\documentclass[]{aa}  

\usepackage{graphicx}
\usepackage{txfonts}
\usepackage[]{hyperref}
\begin{document}

   \title{The SPIRou wavelength calibration for precise radial velocities in the near infrared\thanks{Based on observations obtained at the Canada-France-Hawaii Telescope (CFHT), which is operated from the summit of Maunakea by the National Research Council of Canada, the Institut National des Sciences de l'Univers of the Centre National de la Recherche Scientifique of France, and the University of Hawaii. The observations at the Canada-France-Hawaii Telescope were performed with care and respect from the summit of Maunakea which is a significant cultural and historic site. Based on observations obtained with SPIRou, an international project led by Institut de Recherche en Astrophysique et Planétologie, Toulouse, France.}\fnmsep\thanks{Table A1 is only available in electronic form
at the CDS via anonymous ftp to cdsarc.u-strasbg.fr (130.79.128.5)
or via http://cdsweb.u-strasbg.fr/cgi-bin/qcat?J/A+A/}}

%   \subtitle{bla}

   \author{M. J. Hobson
          \inst{\ref{i:lam}, \ref{i:mia}}%,\ref{i:PUC}, }
          \and
          F. Bouchy\inst{\ref{i:geneve}}
          \and
          N. J. Cook\inst{\ref{i:irex}}
          \and
          E. Artigau\inst{\ref{i:irex}}
          \and        
          C. Moutou\inst{\ref{i:irap1},\ref{i:cfht}}
          \and
          I. Boisse\inst{\ref{i:lam}}
          \and
          C. Lovis\inst{\ref{i:geneve}}
          \and
          A. Carmona\inst{\ref{i:grenoble}}
          \and
          X. Delfosse\inst{\ref{i:grenoble}}
          \and
          J.-F. Donati\inst{\ref{i:irap1}}
          \and
          the SPIRou Team
           }

   \institute{Aix Marseille Univ, CNRS, CNES, LAM, Marseille, France\\
              \email{mhobson@astro.puc.cl}\label{i:lam}
        \and
        Millennium Institute for Astrophysics, Chile\label{i:mia}
        \and
        Observatoire Astronomique de l’Université de Genève, 51 Chemin des Maillettes, 1290 Versoix, Switzerland\label{i:geneve}
        \and
        Institut de Recherche sur les Exoplanètes (IREx), Département de Physique, Université de Montréal, C.P. 6128, Succ. Centre-Ville, Montréal, QC, H3C 3J7, Canada\label{i:irex}
        \and
        Univ. de Toulouse, CNRS, IRAP, 14 avenue Belin, 31400 Toulouse, France\label{i:irap1}
        \and
        Canada-France-Hawaii Telescope Corporation, 65-1238 Mamalahoa Hwy, Kamuela, HI 96743, USA\label{i:cfht}
        \and
        Univ. Grenoble Alpes, CNRS, IPAG, 38000 Grenoble, France\label{i:grenoble}
        }

   \date{Received 13 May 2020, accepted 18 January 2021}

% \abstract{}{}{}{}{} 
% 5 {} token are mandatory
 
  \abstract
   {}{SPIRou is a near-infrared (nIR) spectropolarimeter at the CFHT, covering the YJHK nIR spectral bands ($980-2350\,\mathrm{nm}$). We describe the development and current status of the SPIRou wavelength calibration in order to obtain precise radial velocities (RVs) in the nIR.}{We make use of a UNe hollow-cathode lamp and a Fabry-Pérot étalon to calibrate the pixel-wavelength correspondence for SPIRou. Different methods are developed for identifying the hollow-cathode lines, for calibrating the wavelength dependence of the Fabry-Pérot cavity width, and for combining the two calibrators.}{The hollow-cathode spectra alone do not provide a sufficiently accurate wavelength solution to meet the design requirements of an internal error of $\mathrm{<0.45\,m\,s^{-1}}$, for an overall RV precision of $\mathrm{1\,m\,s^{-1}}$. However, the combination with the Fabry-Pérot spectra allows for significant improvements, leading to an internal error of $\mathrm{\sim 0.15\,m\,s^{-1}}$. We examine the inter-night stability, intra-night stability, and impact on the stellar RVs of the wavelength solution.}{}

   \keywords{Astronomical instrumentation, methods and techniques -- Instrumentation: spectrographs}

   \maketitle
%
%-------------------------------------------------------------------
\section{Introduction}

The spectroscopic method of exoplanet detection, also known as the radial velocity (RV) method, has proven itself extremely productive, with over 700 exoplanets discovered by this method\footnote{See e.g. The Exoplanets Encyclopaedia (\url{http://exoplanet.eu/}), the NASA Exoplanet Archive (\url{https://exoplanetarchive.ipac.caltech.edu/}).}. The detection of exoplanets through spectroscopy relies on the extremely precise measurement of tiny shifts of the stellar spectral lines. In order to make these measurements, a precise wavelength solution is in turn required. 

Historically, two main instrumental approaches have been used for wavelength calibration: iodine cells and hollow-cathode (HC) lamps. Each of these calibrators has advantages and disadvantages. With iodine cells, the calibrating spectrum is imprinted directly on the stellar spectrum, allowing for a fully simultaneous calibration \citep[e.g.][]{Marcy92, Butler96}. However, the iodine lines span only a small portion of the optical spectrum (around 510 to 620 nm, \citealt{Fischer16}), and a high signal-to-noise ratio (S/N) is required to model the line spread function. With HC lamps (primarily ThAr in the visible), calibration exposures must be taken before the stellar observations, and a separate fibre is required to monitor the instrument drift from the time of calibration \citep[e.g.][]{Baranne96}. The advantage of these lamps is that the thorium lines cover a much larger wavelength domain than the iodine lines, and - as they are not superimposed on the stellar spectrum - fainter targets can be observed. An overview of the main spectrographs operating in the visible, their wavelength calibrators, and measurement precisions, is given in \cite{Fischer16}. For iodine cells, the precision is limited to $\mathrm{\approx\,1\,m\,s^{-1}}$ (\citealt{Fischer16}, \citealt{Spronck15}). As for the ThAr HC lamps, \cite{Lovis07} were able to achieve a groundbreaking $\mathrm{20\,cm\,s^{-1}}$ precision for the wavelength solution of the HARPS spectrograph through the creation of an improved line list. For the near-infrared (nIR), however, HC lamps alone limit precision to above $\mathrm{1\,m\,s^{-1}}$ \citep{Halverson14}, and the fill gases emit bright lines that saturate the detectors \citep{Quirrenbach18}.
%TBC refs

In order to increase the precision of the wavelength solution and enable the detection of smaller planets, Fabry-Pérot (FP) étalons and laser frequency combs (LFCs) have recently begun to be incorporated into wavelength solutions. Fabry-Pérot étalons provide lines that are evenly spaced in frequency, but whose wavelengths need to be derived by anchoring to an absolute calibrator, such as an HC lamp \citep[e.g.][]{Bauer15}. As an example, the HARPS spectrograph can achieve $\mathrm{10\,cm\,s^{-1}}$ precision over one night through the incorporation of the FP lines \citep{Wildi11}. Likewise, the CARMENES spectrograph incorporates FP observations into the wavelength calibrations for both its visible and nIR arms; for the nIR arm especially, it is a necessity due to the less densely populated emission lines and strongly saturated gas lines in the HC spectra \citep{Quirrenbach18}. Laser frequency combs, on the other hand, provide evenly spaced (in frequency) lines with wavelengths that are known to very high accuracy \citep[e.g.][]{Murphy07}; their main limitation is the wavelength span they can cover \citep[e.g.][who describe the HARPS LFC, which currently spans around three quarters of the HARPS domain]{Coffinet19}. Some nIR spectrographs currently use LFCs as their primary calibrators, such as the Habitable Zone Planet Finder \citep[HPF][]{Halverson14} and the InfraRed Doppler (IRD) spectrograph \citep{Kokubo16, Kotani18}; however, the wavelength coverage of these spectrographs is smaller than that of SPIRou, not extending as far into the red. Finally, LFCs are also much more costly than FP étalons, and are still maturing as a technology (HPF and IRD both employ HC lamps and FP étalons as back-up calibrators). An LFC covering the 1.0 - 2.2 micron range has recently been installed on SPIRou and is in a testing and optimisation phase, but is not currently part of the standard calibrations \citep{Donati20}.

In this article, we describe the development of the SPIRou wavelength solution, using a UNe HC lamp and an FP étalon. Section \ref{s: data} presents the SPIRou wavelength calibrators and the datasets selected for testing. Section \ref{s: methods} describes the different approaches tested. The performances of the different wavelength solution scripts are analysed in Sect. \ref{s:Perf}. Finally, we summarise and conclude in Sect. \ref{s: Conc}.

\section{Data} \label{s: data}

\subsection{SPIRou spectrograph and calibration unit}

SPIRou (SpectroPolarimetre InfraRouge) is an nIR spectropolarimeter, mounted on the 3.6 m Canada-France-Hawaii Telescope (CFHT), which began operations in February 2019. An overview of the optical and mechanical design is is given in \cite{Artigau}, \cite{Donati18} and \cite{Donati20}\footnote{See also the project website, \url{http://spirou.irap.omp.eu}, and the CFHT instrument page, \url{https://www.cfht.hawaii.edu/Instruments/SPIRou/}}. SPIRou's nominal spectral range, on which the design was optimised, is the $980-2350\,\mathrm{nm}$ wavelength range over 46 echelle orders. In practice, SPIRou covers a total spectral range (with one gap) of $950-2500\,\mathrm{nm}$ (Y, J, H, and K nIR bands) over 50 echelle orders, at $R \approx 70\,000 \pm 3000$ for stellar spectra, using a $4096 \times 4096$ Hawaii 4RG detector. There are three spectral channels, two corresponding to the science observations (in order to simultaneously record the two orthogonal states of a given polarisation state, either circular or linear), and one for calibration.

%The SPIRou calibration unit has been previously described in \cite{Boisse16}, and \cite{Perruchot18}. The calibration module permits the illumination of science, calibration, or all channels simultaneously, by:
%\begin{itemize}
%    \item A cold source (Black Acktar surface at -25 \degr C) for observation of faint stars without simultaneous calibration;
%    \item A white lamp (Tg) for blaze measurement;
%    \item One of two HC lamps, UNe or ThAr, for wavelength calibration;
%    \item An FP \'etalon for wavelength calibration and simultaneous drift monitoring;
%    \item A reserve port intended for future upgrades or visitor instruments.
%\end{itemize}

The SPIRou calibration unit has been previously described in \cite{Boisse16}, and \cite{Perruchot18}. The calibration module permits the illumination of science, calibration, or all channels simultaneously, by any of the following: a cold source (Black Acktar surface at -25 \degr C) for observation of faint stars without simultaneous calibration; a white lamp (Tg) for blaze measurement; one of two HC lamps, UNe or ThAr, for wavelength calibration; an FP \'etalon for wavelength calibration and simultaneous drift monitoring; or a reserve port intended for future upgrades or visitor instruments.

\begin{figure}[htb]
    \centering
    \includegraphics[width=\hsize]{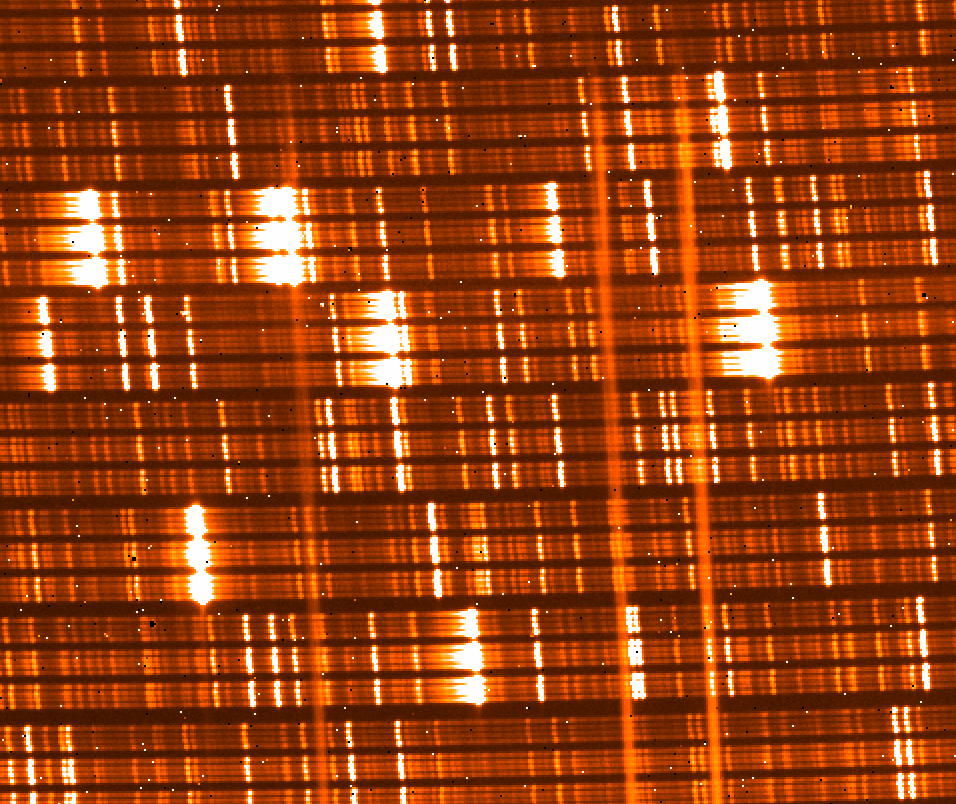}
    \caption{Raw SPIRou spectrum of the UNe HC lamp (zoom to the central region). Light from the lamp is being fed to both the science and calibration fibres. Around eight orders can be seen, centred around 1560 nm. Three 'ghost' lines are also visible, corresponding to contamination from strongly saturated lines in other orders.}
    \label{fig:UNe}
\end{figure}

\begin{figure}[htb]
    \centering
    \includegraphics[width=\hsize]{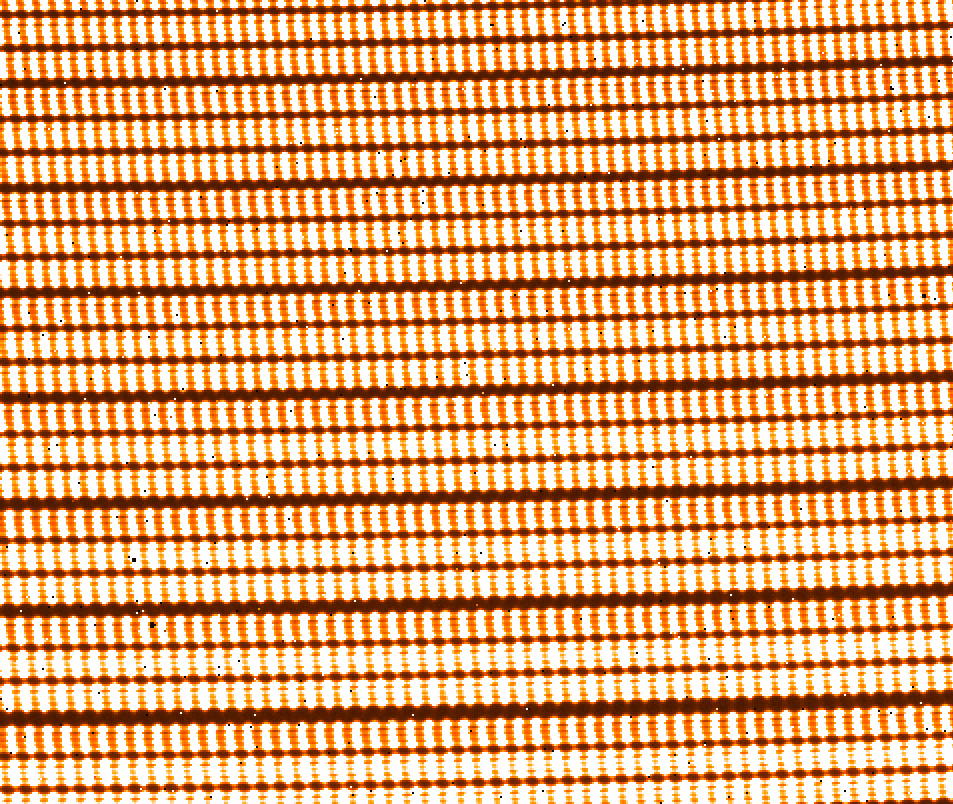}
    \caption{Raw SPIRou spectrum of the FP \'etalon (zoom to the same central region as Fig. \ref{fig:UNe}). Light from the lamp is being fed to both the science and calibration fibres. Around eight orders can be seen, centred around 1560 nm.}
    \label{fig:FP}
\end{figure}

The wavelength calibrators are the HC lamps and the FP étalon. Figures \ref{fig:UNe} and \ref{fig:FP} show the central regions of raw SPIRou spectra of the UNe lamp and the FP \'etalon, respectively. Each of these calibrators has advantages and disadvantages. Either of the HC lamps can provide an absolute calibration, through identifiable catalogued lines; however, the lines are unevenly spaced and vary significantly in flux. Meanwhile, FP lines do not have fixed absolute wavelengths but must be anchored to another calibrator; however, once a first absolute calibration is obtained from the HC lamp, the multitude of evenly spaced FP lines across the entire detector enable a refinement of the wavelength solution. 

SPIRou aims for a target RV precision of $\mathrm{1\,m\,s^{-1}}$. In order to achieve this precision, the error budget requires for the wavelength solution an internal error of $\mathrm{<0.45\,m\,s^{-1}}$.

\subsection{Hollow-cathode lamp selection}

\begin{figure}
\centering
\includegraphics[width=\hsize]{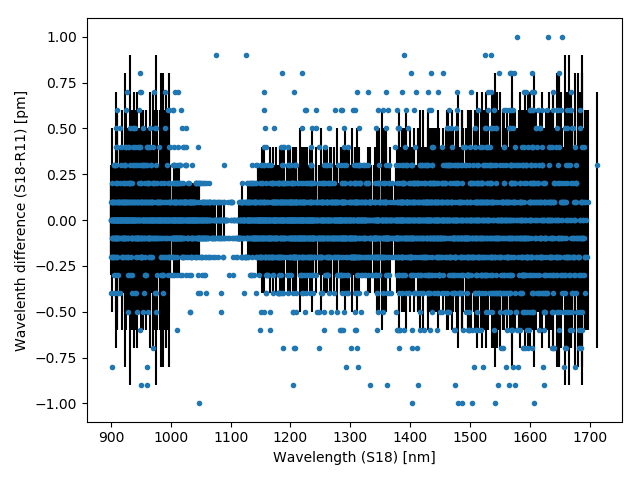}
\caption[UNe catalogues comparison]{Comparison of wavelengths for common lines between the catalogues of \cite{Redman11} (R11) and \cite{Sarmiento18} (S18). The blue points represent the difference in wavelengths, the black lines the S18 error bars. Points without error bars indicate that S18 reported no uncertainty for that line. The differences in wavelength are generally within the error bars when known, and always small.}
\label{fig:cat-comp}
\end{figure}

At the start of SPIRou commissioning, the only available catalogue of UNe lines in the nIR was that of \cite{Redman11} (hereafter R11), covering the $850-4000\,\mathrm{nm}$ wavelength range, which fully encompasses the SPIRou wavelength range of $980-2350\,\mathrm{nm}$. In particular, it has 9767 lines in the SPIRou wavelength range (median uncertainty: 0.2 pm, translating to a median RV uncertainty per line of 35 $\mathrm{m\,s^{-1}}$). This catalogue was therefore used in the initial development of the HC wavelength solution during validation tests. In 2018, however, \cite{Sarmiento18} (hereafter S18) published an updated catalogue of U lines in the $500-1700\,\mathrm{nm}$ wavelength range. While this is far from covering the full wavelength range of SPIRou, encompassing only the bluer half, it cross-matches well with the catalogue of R11 (Fig. \ref{fig:cat-comp}) and provides a substantial increase in the range it does cover, incorporating 3787 new lines (median uncertainty of all lines in the catalogue: 0.9 pm, translating to a median RV uncertainty per line of 29 $\mathrm{m\,s^{-1}}$) for a total of 13554 lines in a combined R11+S18 catalogue. 

For the ThAr lamp, the only catalogue that covered the SPIRou domain at the time was that of \cite{Redman14}  (median uncertainty: 0.12 pm, translating to a median RV uncertainty per line of 7 $\mathrm{m\,s^{-1}}$). The UNe catalogue has far more lines than the ThAr catalogue, even before the incorporation of the lines from S18 (9767 vs 1587 in the SPIRou wavelength range).

The laboratory tests performed at Toulouse, which are described in \cite{Perruchot18}, show that the UNe spectra have around four times more identifiable lines than the ThAr spectra, resulting in consistently more accurate and stable wavelength solutions throughout the tests. Additionally, the ThAr spectra show many more, and more strongly, saturated lines than the UNe spectra in the SPIRou wavelength range. Saturated lines are useless for fitting a wavelength solution, since their centres cannot be precisely determined. They will also 'bleed' into neighbouring orders and contaminate them. Additionally, persistence, which is a known problem for CMOS detectors such as the H4RG \citep{Artigau18, Bechter19}, is stronger for saturated lines, contaminating subsequent observations. Therefore, the UNe lamp was adopted as the primary absolute wavelength calibrator.

\subsection{Selected datasets for testing}

To showcase the performance of the different methods developed, we applied all the scripts on the calibrations of a two-week SPIRou run in February 2019\footnote{All SPIRou calibrations are publicly available via the Canadian Astronomy Data Centre, at \url{http://www.cadc-ccda.hia-iha.nrc-cnrc.gc.ca/en/}}. The selected calibrations consist of pairs of 1 HC (UNe) spectrum and 1 FP spectrum. The many instrument changes over the comissioning and early science runs prevent us from making longer-term comparisons.

\section{Methods} \label{s: methods}

\subsection{SPIRou data reduction system}

The wavelength solution is derived for each night as part of APERO (A PipelinE to Reduce Observations); APERO is maintained and version-controlled on github \footnote{Located at \url{https://github.com/njcuk9999/apero-drs}}.
For the remainder of this work we will refer to the SPIRou specific part of APERO as the SPIRou data reduction system (DRS). The work presented here was carried out primarily with version 0.5.000, which was released on 10 May 2019. The main differences with the current development version are noted in Sect. \ref{s: Conc}, and will be described in a forthcoming paper (N.J. Cook et al. in prep).

A full description of the DRS is beyond the scope of this work, and will be done in N.J. Cook et al. (in prep). However, we briefly summarise the calibration processing sequence for version 0.5.000: pre-processing to remove certain detector effects such as the amplificator crosstalk; generation of a dark calibration; creation of a map of the bad pixels, including two small holes and a scratch on the detector; localisation of the orders for fibres A, B, and C;  mapping of the slit profile across the detector; creation of the blaze profiles for all extracted fibres; generation of the wavelength map for all extracted fibres.

\subsection{Wavelength solution using HC spectra alone} \label{HC-sol-desc}

Two main methods were developed and tested for the wavelength solution based on the HC lamp. The principal conceptual difference between them lies in the way in which the HC lines are identified. The first, method HC1, is analogous to the SOPHIE/HARPS wavelength solution \citep{Baranne96}: For each line in the catalogue, the region where it should be located is selected and a Gaussian fit is attempted, with poor fits being discarded. In the second, method HC2, Gaussians are fitted to every peak in the HC spectra, and the best match to the catalogue is identified. The detailed structure of the full routine is as follows.

\paragraph{Data identification and reading:} The input files are verified via FITS header keys to be extracted two-dimensional spectra (E2DS) corresponding to an HC lamp. The E2DS files contain the extracted spectrum for a single fibre in the format of a $\mathrm{50 \times 4089}$ array with one spectral order per row, arranged from blue to red. The data and header are read, and the lamp and fibre are identified. If more than one HC file is given, the routine verifies they all correspond to the same lamp and fibre, then the median of the frames is obtained and used as input for the rest of the routine.

\paragraph{Calibration set-up:} The calibration files that are closest in time to the input file(s) are copied from the calibration database (including the previous wavelength solution). The previous wavelength solution is read and checked for compatibility with the current parameter set-up (degree of the polynomial fits employed, as this varied during development). The correct HC line catalogue (UNe or ThAr) is read in.

\paragraph{Identification of the HC lines:} The lines are identified following one of the procedures outlined above:

\textit{Method HC1:} For each line in the catalogue, a $4 \sigma$ region in wavelength (where $\sigma$ is the expected width of the line, computed from the central wavelength and the average spectral resolution) around its expected position is selected, based on the previous wavelength solution. A Gaussian fit to the region is attempted, and kept if the $\sigma$ of the fitted Gaussian is less than $\mathrm{4\,km\,s^{-1}}$ (to discard very broad, flat lines whose centres will be imprecise), and the amplitude is less than 1e8 (to avoid fitting on saturated lines; we note that the amplitude is dimensionless as the E2DS files are blaze-corrected within the routine).

\textit{Method HC2:} For each spectral order, a 13-pixel-wide window is moved along the order in four-pixel shifts. If the maximum flux is at least at a $2 \sigma$ level above the local RMS and is within four pixels of the centre of the segment, a Gaussian fit is attempted. The Gaussian is kept if the residual of the fit normalised by the peak value is between 0 and 0.2 (to ensure the fit is a good representation of the spectrum in that region), and the Gaussian $\sigma$ in pixels, assuming a FWHM of 2 pixels, is between 0.7 and 1.1 (to discard narrow cosmic rays or broad blended lines). Once all the peaks in the order are identified, the 20 brightest (which are generally the least likely to be spurious, and most likely to correspond to catalogued lines) are selected and the closest catalogue line to each one is identified. Then, all possible three-line combinations are used to fit a second-order polynomial to the order, test wavelengths are calculated for all lines using the polynomial fit, and the velocity offset for each from its identified catalogue line is calculated. The polynomial with the most lines within $\mathrm{1\,km\,s^{-1}}$ of the catalogue is held to be the correct identification, and all peaks within $\mathrm{1\,km\,s^{-1}}$ of the catalogue are kept.

The parameter values were set and refined over the course of the pipeline development in order to obtain the most accurate and stable wavelength solutions. For method HC1 in particular, the SOPHIE/HARPS parameters were used as first guesses, and modified as necessary. The fitted Gaussians to the HC lines have a median FWHM of $\mathrm{2.1}$ pixels, corresponding to wavelength values ranging from $\mathrm{0.01\,nm}$ at the blue end of the spectrum to $\mathrm{0.05\,nm}$ at the red end.

\paragraph{Fitting the solution:} With the HC lines identified, a fourth-order polynomial fit is performed for each order between the pixel positions given by the centres of the Gaussians, and the catalogued wavelengths. In method HC2, continuity of the polynomial coefficients across the orders is also imposed at this step (attempts to impose similar cross-order continuity for method HC1 resulted in less stable solutions, so they were discarded).

\begin{figure}[htb]
	\centering
	\includegraphics[width=1\linewidth]{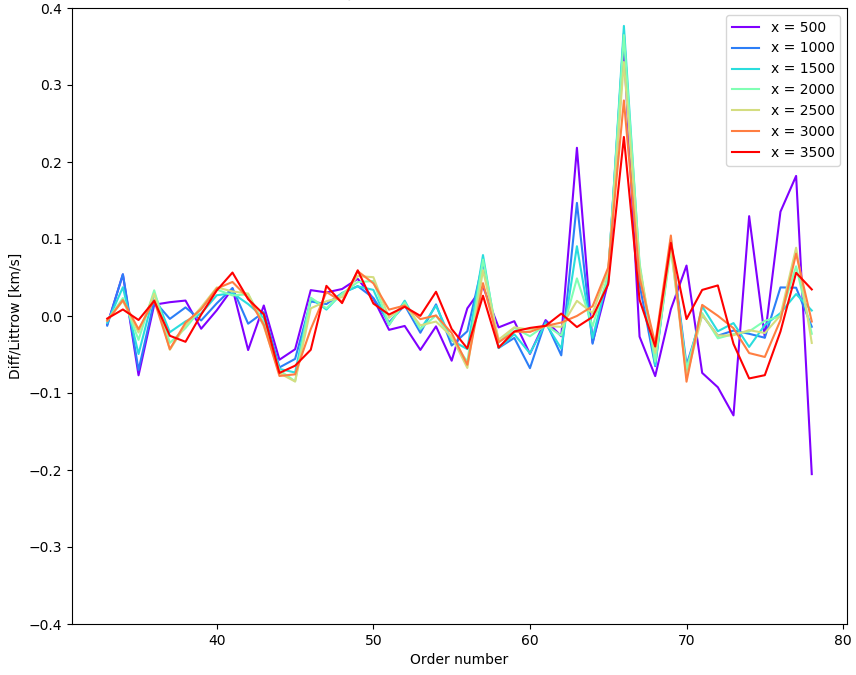}  
\caption[Littrow check]{Example Littrow solution check, showing the residuals of cross-order fits between the inverse echelle orders and the fractional wavelength contribution. Each line corresponds to a different pixel position. It should be noted that for the bluest orders, there is very little flux at pixels 500 and 3500.}
	\label{fig: Littrow}
\end{figure}

\paragraph{Littrow solution check:} The Littrow check is a verification of the cross-order continuity of the wavelength solution, following the same principles as in \cite{Cersullo19}. It is evaluated for several pixel positions (currently every 500 pixels along each order in the dispersion direction). For each position x, a fourth-order polynomial fit is performed between the inverse echelle orders and the fractional wavelength contribution at x for each order (normalised by the wavelength for the first order), with an iterative step to remove the largest outlier. For an ideal spectrograph, the residuals to these fits would be zero. The polynomial fits, the residuals, and the minimum, maximum, and rms values of the residuals are stored. An example is shown in Fig. \ref{fig: Littrow}.

\paragraph{Extrapolation of the reddest orders:} For the last two orders, very few HC lines are catalogued (13 in the last order, wavelength range 2438-2516 nm, and 68 in the penultimate, wavelength range 2362-2437 nm, compared to an average of 300 for the rest) and even fewer are identified (around ten in the last order and 25 in the second-to-last for method HC2). This means that the fitted solution often fails or is highly unstable. Therefore, for these orders it is not fitted from the HC lines, but extrapolated from the Littrow solution: The cross-order polynomial fits are used to generate pixel-wavelength pairs at the Littrow evaluation positions, and these values are used in turn to fit a fourth-order polynomial for each spectral order. A possible alternative would be to use telluric lines to fit these two orders; however, work by \cite{Figueira10} using CRIRES suggests the precision would not be better than $\mathrm{5\textnormal{-}10\,m\,s^{-1}}$.

\paragraph{Quality controls:} The structure of an E2DS file (one order per row, wavelength increasing along the order) means that along each column, the wavelength must be increasing - that is, each pixel of order N must have a smaller wavelength value than the corresponding pixel of order N+1. A first rough quality check, therefore, verifies that this is in fact the case (when this fails, it generally points to a problem with the slit shape or order identification calibrations). A second quality check is applied to the minimum, maximum, and rms values of the Littrow check residuals: The minimum and maximum must be within $\mathrm{0.3\,km\,s^{-1}}$, the rms within $\mathrm{0.1\,km\,s^{-1}}$.

\paragraph{Logging statistics:} The mean and rms of the deviation from the catalogued lines in $\mathrm{m\,s^{-1}}$ is logged, as is the total number of lines used to fit the solution. Most importantly, the internal precision of the solution is determined, as the rms of the residuals of the fit to the catalogued lines, divided by the number of lines used to generate the fit.

\paragraph{Saving the solution:} The wavelengths per pixel generated from the polynomial fits for each order are saved as an E2DS file. The coefficients of the polynomial fits are stored in the header. If the quality controls were all passed, the new wavelength solution is copied to the calibration database, and to the header of the input HC E2DS spectra.

Performance tests (discussed in more detail in Sect. \ref{s:perf-HC}) showed that while the first method has slightly better internal accuracy than the second, it is much less stable night-to-night. In any case, neither method reaches the expected internal accuracy of $\mathrm{<0.45\,m\,s^{-1}}$ budgeted for the SPIRou wavelength solution. The first method is also very sensitive to drifts from the initial wavelength solution (this was particularly highlighted by the earthquake in May 2018, which caused a detector shift of several pixels), to which the second method is more robust. The second method was therefore adopted for the HC wavelength solution, and is the only one available in version 0.5.000 of the DRS.

During the SPIRou validation and commissioning tests, it became clear that the HC lamps alone did not provide a sufficiently accurate and stable wavelength solution, with internal accuracy measurements of $\mathrm{\sim 2{\text -}4\,m\,s^{-1}}$ compared to the $\mathrm{0.45\,m\,s^{-1}}$ accuracy demanded by the SPIRou error budget for an overall $\mathrm{1\,m\,s^{-1}}$ RV precision. There are likely several contributing factors to this lack of accuracy: the low flux levels in the edges of the bluest orders, the low number of lines found for the reddest orders, imprecision in the catalogue wavelengths, among others. The next step, therefore, was to combine the HC lamps with the FP étalon spectra. 

\subsection{Wavelength solution combining HC and FP spectra} \label{FP-sol-desc}

The design of the SPIRou FP étalon is described in \cite{Cersullo17}. It has a finesse value of F = 12.8, and was designed to cover the entire $980-2350\,\mathrm{nm}$ wavelength range of SPIRou. Its spectrum provides a wealth of lines across the entire detector, as shown in Fig. \ref{fig:FP}, whose spacing is a priori known since it is given by the FP equation (though we will see that for a physical as opposed to an ideal étalon, this is not entirely accurate as the cavity width is wavelength-dependent). The absolute wavelengths of the FP lines, however, are not known but must be determined from some other source. Therefore, by anchoring the FP line wavelengths to the HC lines, we aimed to derive a more precise wavelength solution than could be obtained with the HC alone.

In principle, the wavelength of an FP line is given by the FP equation:
\begin{equation}\label{FPeq}
    \lambda_m = \frac{2d}{m},
\end{equation}
where $\lambda_m$ is the wavelength of the line of (integer) line number $m$, and $d$ is the effective cavity width of the interferometer. In an ideal FP interferometer, $d$ is constant (and known since it is set by the manufacturer); therefore, once the wavelength $\lambda_{m,r}$ is known for a specific reference line, its line number $m_r$ can be determined. Then, for any other line, the line number can be determined simply by counting from the reference line, and its wavelength calculated with Eq. \ref{FPeq}. However, in real FP \'etalons, the cavity width $d$ is not constant. \cite{Bauer15} showed it to be wavelength-dependent, due to a varying penetration depth; photons of different energy (i.e. different wavelength) will penetrate the soft coating to different depths. This wavelength dependence needs to be calibrated in order to allow the use of the FP lines in a wavelength solution.

Here, again, two methods were developed. The first, method FP1, follows the approach described by \cite{Bauer15}: We first use the HC lines to generate a rough wavelength solution, from which we obtain first-guess FP wavelengths; these first-guess wavelengths are in turn used to fit the cavity width. The second, method FP2, is based on the one developed by C. Lovis for ESPRESSO (private communication). Fractional FP line numbers are assigned to the HC lines, which are then used to fit the cavity width directly. The overall structure of the algorithm is as follows.

\paragraph{Data identification and reading:} The input files are verified via FITS header keys to be E2DS files corresponding to an HC lamp and the FP étalon, respectively. The data and header are read, and the lamp and fibre are identified. Fibre correspondence between the HC and FP files is checked. If more than one HC file is given, the median of the frames is obtained and used as input for the rest of the routine. Currently, providing more than one FP file is not supported. This option was chosen as allowing for multiples of each file type increased set-up complexity, and the FP spectra are generally bright everywhere while the HC lines can be weaker.

\paragraph{Calibration set-up:} Previous calibration files are copied from the calibration database (including the blaze file and the previous wavelength solution). The previous wavelength solution is read and checked for compatibility with the current parameter set-up (order of the polynomial fits employed). The correct HC line catalogue is read in.

\paragraph{Generation of a first-guess wavelength solution:} The HC spectra are used to generate a wavelength solution, using Method HC2 from Sect. \ref{HC-sol-desc}. Quality controls are applied to this first-guess solution. 

\paragraph{Incorporation of the FP lines:} The FP lines are identified and Gaussians fitted to them: For each order, the highest value is identified, and a Gaussian fit attempted on a 7-pixel box around it. Fits that do not fail and are centred within $\pm 1$ pixel of the pixel with the highest value are stored, the rest are rejected. Finally the Gaussian fit is subtracted (or the region set to zero if the fit fails), and the process iterates until no more lines are found. The Gaussians fitted to the FP have a median FWHM of $\mathrm{2.4}$ pixels, corresponding to wavelength values ranging from $\mathrm{0.01\,nm}$ at the blue end of the spectrum to $\mathrm{0.05\,nm}$ at the red end.  With the FP lines' pixel positions known, the initial wavelength solution generated from the HC spectra alone is used to fit first-guess FP wavelengths from the FP line pixel positions. Using the FP equation and an input cavity width value, the FP line number is obtained for the last peak of the reddest order. We initially used the manufacturer's value of $\mathrm{d = 12.25\,mm}$ before coating, meaning $\mathrm{2d = 24.5\,mm}$, as the input value; subsequent testing allowed us to refine it to $\mathrm{2d = 24.4999\,mm}$. The rest of the lines are then numbered by counting along each order, using wavelength matching across orders (with gaps due to missed peaks accounted for). It does not matter whether the same last peak is found for each observation, as we find that the variation of the cavity width within the order is small enough that the same FP line numbers are obtained from the FP equation using the input cavity width value as by counting from the last peak. 

\begin{figure}[htb]
	\centering
	\includegraphics[width=1\linewidth]{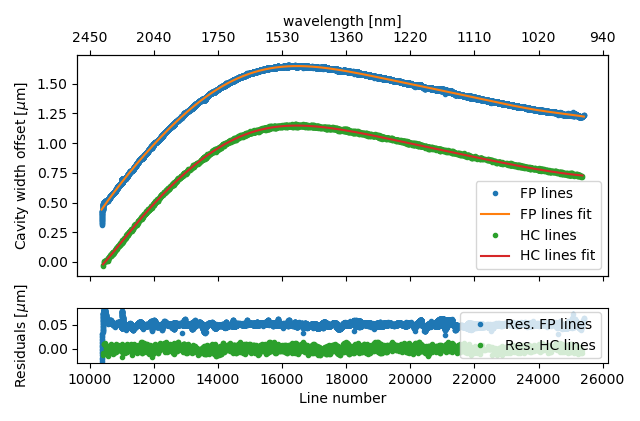}  
	\caption[Variability of the FP cavity width]{Variability of the FP cavity width with regard to the input value for methods FP1 (FP lines) and FP2 (HC lines), offset by 0.5 $\mathrm{\mu}$m for visibility (top), and residuals to the fits, offset by 0.05 $\mathrm{\mu}$m for visibility (bottom). The results from the two methods are very similar. For method FP1, some outliers (poorly fitted FP lines) can be seen, while for method FP2 the number of lines drops off towards low line numbers (i.e. high wavelengths) and no lines are selected for the reddest order. For scale, a cavity width offset of 0.25 $\mathrm{\mu}$m corresponds to a velocity offset of $\mathrm{6.12\,km\,s^{-1}}$, using the FP equation and the initial cavity width value of $\mathrm{2d = 24.4999\,mm}$.}
	\label{fig: FP-cavity-width}
\end{figure}

\paragraph{Wavelength dependence of the cavity width:} With the FP line numbers identified, the wavelength dependence of the cavity width is dealt with using one of the two methods named above:

\textit{Method FP1:} Using the line numbers and the first-guess wavelengths of the FP lines (derived from the first-guess HC solution previously generated), a cavity width is calculated for each peak. A ninth-order polynomial is fitted to the cavity width (Fig. \ref{fig: FP-cavity-width}, top points and fit), and corrected FP line wavelengths are calculated from this polynomial fit. 

\textit{Method FP2:} First, a fourth-order polynomial is fitted to the FP line numbers as a function of pixel positions per order. This fit is used to generate fractional line numbers for the 'best' HC lines (selected as the lines at blaze values of more than 30\%, and with velocity offset from the catalogue less than $\mathrm{0.25\,km\,s^{-1}}$). Using these fractional line numbers and the catalogue wavelengths, the cavity width is calculated for each HC line using the FP equation. Ninth-order polynomials are then fitted to these cavity width values (Fig. \ref{fig: FP-cavity-width}, bottom points and fit) as a function of both line number and wavelength, and corrected FP line wavelengths are calculated from the fit. Optionally, a previous cavity width fit can be read in; in this case we assume only an achromatic change (i.e. a shift) may have taken place, and correct the fit for this shift by subtracting the median of the residuals between the newly calculated cavity widths for the HC lines and the previous fit.

\paragraph{Fitting the solution:} The HC and FP lines are combined. A fourth-order polynomial fit is performed between the pixel positions, given by the centres of the Gaussians, and the line wavelengths (catalogued for the HC lines, generated from the cavity width fit for the FP lines).

\paragraph{Calculating the FP RV:} The RV of the FP spectrum is calculated using the cross-correlation function (CCF) method, through cross-correlation with an FP mask. It is stored in order to give an FP RV zero-point, from which the drift of the spectrograph for a later observation of a star with simultaneous FP can be measured (by subtracting the RV of the wavelength solution's FP from the RV of the simultaneous FP). Currently, we assume the drift between the HC exposure that gives the absolute zero-point and an immediately subsequent FP exposure to be negligible, given the stability of SPIRou. In the future, a possible next step would be to employ HC-FP and FP-HC exposures (i.e. exposures with the HC lamp on the science fibres and the FP étalon on the calibration fibre or vice versa) to calibrate this drift, as done in, for example, ESPRESSO or HARPS.  

\paragraph{Littrow solution check, quality controls, logging statistics:} These three steps are analogous to those of the HC solution.

\paragraph{Saving the solution:} Similar to the HC solution. If the quality controls were all passed, the header of the input FP E2DS spectra is also updated. The HC and HC-FP solutions are stored under different file names, so that the HC-FP solution does not overwrite the HC solution.

\paragraph{Saving results tables:} Two tables are stored. The first logs the statistics of the Littrow solution quality check. The second is a list of all lines used for the solution, containing the order, wavelength, difference in velocity of the final fit from the input line value, weight, and pixel position for each line.

% \begin{figure}[htb]
% 	\centering
% 	\includegraphics[width=1\linewidth]{cavity_width_fit_FP1.png}  
%     \includegraphics[width=1\linewidth]{cavity_width_fit_FP2.png}      
% 	\caption[Variability of the FP cavity width]{Variability of the FP cavity width for method FP1 (top) and method FP2 (bottom), as plotted by the pipeline. The results from the two methods are very similar. For method FP1 some outliers (poorly fitted FP lines) can be seen, while for method FP2 the number of lines drops off towards low line numbers (i.e. high wavelengths) and no lines are selected for the reddest order.}
% 	\label{fig: FP-cavity-width}
% \end{figure}

As will be discussed in Sect. \ref{s:perf-HC-FP}, the two methods are comparable, though method FP2 has slightly better accuracy and stability. In version 0.5.000 of the DRS, only method FP1 was available for general use as method FP2 was still in development. In forthcoming versions both will be offered as options in the final wavelength solution algorithm.

\section{Performance tests and validation} \label{s:Perf}

\begin{figure}[htb]
	\centering
    \includegraphics[width=1\hsize]{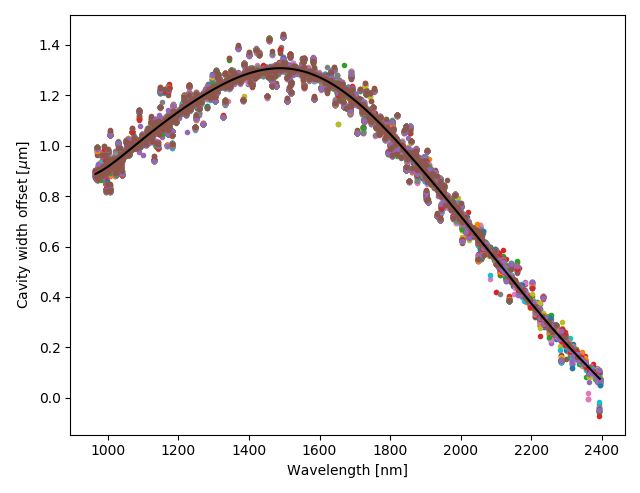}  
	\includegraphics[width=1\hsize]{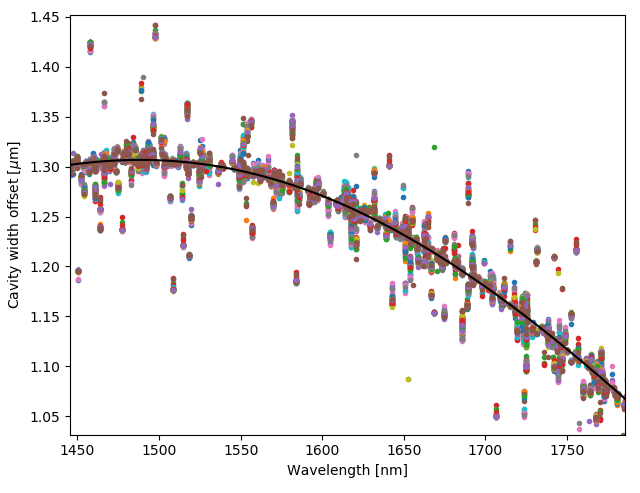}      
	\caption[Construction of a stable lines catalogue]{Construction of a stable lines catalogue. Top: Overall cavity width fit (black line) from the HC lines (dots) for fifty exposures. Bottom: Zoom showing how the values for each line cluster together, but can be offset from the main fit. For scale, a cavity width offset of 0.2 $\mathrm{\mu}$m corresponds to a velocity offset of $\mathrm{4.88\,km\,s^{-1}}$, using the FP equation and the initial cavity width value of $\mathrm{2d = 24.4999\,mm}$.}
	\label{fig: stable-fit-zoom}
\end{figure}

To test the performance of the different wavelength solution generation methods, we ran all scripts on the calibrations of a two-week SPIRou run in February 2019. We used three different UNe line catalogues: the R11 catalogue, a combination of the R11 and S18 catalogues (with the S18 wavelengths kept for matching lines), and a selection of the most stable lines (i.e. those consistently identified for different HC frames) with updated, more accurate wavelength values. This selection of lines was derived using method FP2. First, the cavity width was fitted from the best HC lines (at blaze values of more than 30\%, and with velocity offset from the catalogue less than $\mathrm{0.25\,km\,s^{-1}}$) for each of 50 HC exposures taken during commissioning (between May and November 2018). For each of these exposures, the wavelengths, fractional line numbers and calculated cavity widths of the lines used for the fit were saved. Then, all the lines were combined, and their wavelengths and cavity widths were fitted together to generate a very accurate cavity width fit (Fig \ref{fig: stable-fit-zoom}, top panel). Using this accurate cavity width fit, it can be seen that while for each catalogue line the different cavity widths measured for each exposure cluster together, these clusters can be significantly offset from the overall fit (Fig \ref{fig: stable-fit-zoom}, bottom panel). This would mean that for these offset values, the catalogue wavelengths are inaccurate. Therefore, each measured line's wavelength was recalculated from its fractional line number and the FP equation. Finally, each line that was selected in at least two exposures was assigned an updated wavelength, as the median of its recalculated wavelengths. The stable lines catalogue is therefore not just a selection of the best lines from the others, but a new catalogue with updated wavelength values for each line. The full updated catalogue is presented in Appendix \ref{appendix}.

\subsection{Impact of previous calibrations}\label{ch2:prev-calib-impact}

The wavelength solution is the last step of a long calibration sequence. This made its development particularly challenging as changes upstream frequently meant large modifications in the data that proved destabilising for the wavelength solution, especially in the earliest versions of the DRS. As the DRS has evolved, this has somewhat ameliorated. Nevertheless, the previous calibrations can still impact the wavelength solution. In order to explore the performance of the wavelength solutions alone, therefore, the tests presented in the rest of this section are all carried out with the HC and FP files extracted using a single set of input calibrations. We found that fixing the input calibrations introduces large drifts between the wavelength solutions for different nights, of the order of $\mathrm{\sim 10{\text -}50\,m\,s^{-1}}$; however, they are easily calibrated out by subtracting the night-to-night median RV difference. This removes sensitivity to long-term drifts, without affecting the analysis of the wavelength solution.

\subsection{Performances of the HC solutions} \label{s:perf-HC}

To test the performance of the HC wavelength solutions, we ran both methods on the UNe spectra taken as part of the afternoon calibrations for the two-week SPIRou run in February 2019, for all three wavelength catalogues. The spectra were reduced with a single set of calibrations. We obtained one solution per night for each method and catalogue. Table \ref{tab: HC-sol-perf} summarises the results. The 'internal error' row is a median of the internal accuracies reported for the solutions obtained with the corresponding method and catalogue. The 'local night-to-night variation' represents the median difference between consecutive nights' solutions (computed as the median of the pixel-by-pixel absolute drift-corrected RV differences, with the drift corrected by subtracting the overall median), and can be thought of as the typical uncertainty of a single spectral line. The 'HC lines used' row is the median of the number of HC lines that were identified and used to fit each wavelength solution. The line count is performed per order, so any lines used for more than one order will be counted twice.

\begin{table*}[ht]
    \centering
        \caption[Summary of HC wavelength solution performances.]{Summary of HC wavelength solution performances.}  
    \begin{tabular}{llll}
        \hline \hline
                                           &                          & Method HC1 & Method HC2 \\
        \hline
        R11 catalogue     & Global internal error             & $\mathrm{1.88\,m\,s^{-1}}$   & $\mathrm{3.87\,m\,s^{-1}}$    \\
                                           & Local night-to-night variation & $\mathrm{16.4\,m\,s^{-1}}$ & $\mathrm{6.3\,m\,s^{-1}}$ \\
                                           & HC lines used               & 5607       & 4770       \\
        \hline
        R11+S18 catalogue & Global internal error             & $\mathrm{1.88\,m\,s^{-1}}$    & $\mathrm{3.95\,m\,s^{-1}}$    \\
                                           & Local night-to-night variation & $\mathrm{15.1\,m\,s^{-1}}$ & $\mathrm{7.6\,m\,s^{-1}}$    \\
                                           & HC lines used               & 6186       & 5310       \\
        \hline
        Selected lines      & Global internal error             & $\mathrm{1.46\,m\,s^{-1}}$   & $\mathrm{2.21\,m\,s^{-1}}$    \\
                                           & Local night-to night variation & $\mathrm{13.9\,m\,s^{-1}}$  & $\mathrm{5.7\,m\,s^{-1}}$ \\
                                           & HC lines used               & 2363       & 2124   
                                           \\
        \hline
    \end{tabular}
        \label{tab: HC-sol-perf}
\end{table*}

Method HC1 has better internal accuracy than method HC2, but is less stable from one night to the next. As noted in Sect. \ref{HC-sol-desc}, its sensitivity to the input wavelength solution was particularly highlighted by the earthquakes suffered by the CFHT during SPIRou validation, on 3 and 4 May 2018. The multi-pixel displacement meant all lines were significantly shifted with regard to the search windows defined from the previous (pre-earthquake) solutions. This required the creation of additional algorithms to identify the pixel shifts and generate shifted first-guess solutions, in order for method HC1 to be able to run.  Concern over this sensitivity was in fact one of the driving motivations for the development of method HC2, which (since it identifies all peaks in the spectrum and then generates a best match to the catalogue) is more robust to such shifts. 

Regarding the catalogues, adding the lines from S18 does not seem to create a substantial change in accuracy or stability. Most of the lines added have fairly low relative intensities reported by S18, so they are likely small and their Gaussian fits may be less precise. The 'stable lines' catalogue provides somewhat more accurate and stable wavelength solutions for both methods. In any case, neither method reaches the required internal precision of $\mathrm{0.45\,m\,s^{-1}}$ with any catalogue.

\subsection{Performances of the HC+FP solutions} \label{s:perf-HC-FP}

We used the same two-week SPIRou run in February 2019, processed with a single set of calibrations, to test the performances of both the combined HC-FP wavelength solution methods. To generate the first-guess HC solution, we applied method HC2 in both cases. Once again, we tested the three different wavelength catalogues. Table \ref{tab: FP-sol-perf} summarises the results, with the same rows as for Table \ref{tab: HC-sol-perf}.

\begin{table*}[ht]
    \centering
        \caption[Summary of HC-FP wavelength solution performances.]{Summary of HC-FP wavelength solution performances.}  
    \begin{tabular}{llll}
        \hline \hline
                                           &                          & Method FP1 & Method FP2 \\
        \hline
        R11 catalogue     & Global internal error             & $\mathrm{0.18\,m\,s^{-1}}$    & $\mathrm{0.12\,m\,s^{-1}}$    \\
                                           & Local night-to night variation & $\mathrm{3.1\,m\,s^{-1}}$  & $\mathrm{1.5\,m\,s^{-1}}$ \\
                                           & HC+FP lines used               & 23612       & 21662       \\
        \hline
        R11+S18 catalogue & Global internal error             & $\mathrm{0.18\,m\,s^{-1}}$    & $\mathrm{0.12\,m\,s^{-1}}$    \\
                                           & Local night-to night variation & $\mathrm{3.0\,m\,s^{-1}}$ & $\mathrm{1.9\,m\,s^{-1}}$ \\
                                           & HC+FP lines used               & 23691       & 21874       \\
        \hline
        Selected lines      & Global internal error             & $\mathrm{0.18\,m\,s^{-1}}$    & $\mathrm{0.13\,m\,s^{-1}}$   \\
                                           & Local night-to night variation & $\mathrm{4.3\,m\,s^{-1}}$   & $\mathrm{3.3\,m\,s^{-1}}$    \\
                                           & HC+FP lines used               & 22990       & 20466   
                                           \\
        \hline
        Fixed cavity width    & Global internal error          & NA    & $\mathrm{0.14\,m\,s^{-1}}$     \\
                                           & Local night-to night variation & NA   & $\mathrm{0.8\,m\,s^{-1}}$    \\
        \hline
    \end{tabular}
        \label{tab: FP-sol-perf}
\end{table*}

In this case, the two methods are very comparable, though method FP2 has somewhat better internal accuracy and stability. Rather perplexingly, for the combined HC-FP solutions the 'stable lines' catalogue provides the least night-to-night stability! This is particularly evident for method FP2, where the HC lines are used to fit the cavity width directly. Since this catalogue was generated from a cavity width fit using multiple HC exposures, each reduced with the corresponding nightly calibrations, this may perhaps be a derived effect of the previous calibrations' instability. 
Nevertheless, in all cases the internal accuracy is excellent, and the night-to-night stability is much improved compared to the HC solutions.

As was described in Sect. \ref{FP-sol-desc}, for method FP2 there is an option to read in a previous cavity width fit and correct it from any achromatic shift, instead of generating it anew. The reasoning behind this is that the chromatic dependence is an intrinsic property of the soft coating; while it may evolve slowly over the lifetime of the instrument, it is not expected to change from one night to the next. An achromatic shift, meanwhile, would correspond to a change in the physical separation of the FP, which could be caused by pressure or temperature changes.

\begin{figure}
\centering
\includegraphics[width=1\hsize]{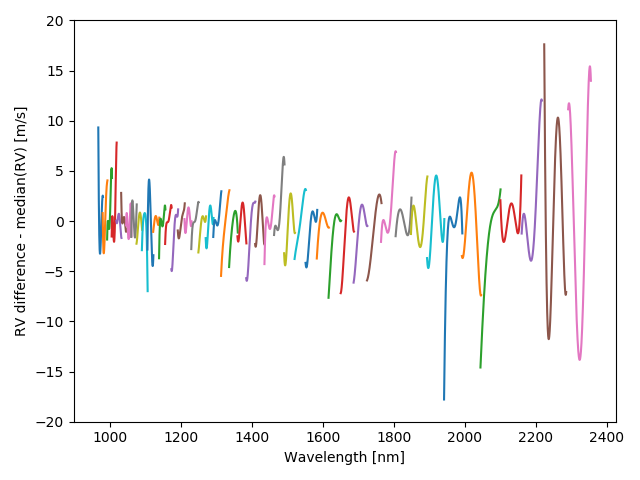}
\caption[Night-to-night variations of the FP solution]{Night-to-night variations of the FP wavelength solution: difference (in RV space) between the solutions generated with method FP2 for the 21 and 22 February 2019, using a fixed cavity width fit. The median value of the  absolute RV differences is $\mathrm{0.8\,m\,s^{-1}}$. The drift induced by the fixed set of prior calibrations has been corrected.}
\label{fig:FP-night-night}
\end{figure}

We tested the implementation of this option, redoing the analysis with an initial cavity width read in. We found a median internal error of $\mathrm{0.14\,m\,s^{-1}}$, and a median night-to-night variation of $\mathrm{0.8\,m\,s^{-1}}$, regardless of the catalogue used. This implies that a substantial part of the night-to-night variability for the combined HC-FP solutions is in fact coming from the cavity width fit. An example of a night-to-night comparison is shown in Fig. \ref{fig:FP-night-night}, for the solutions computed with method FP2 using a fixed cavity width, for the nights of $\mathrm{21^{st}}$ and $\mathrm{22^{nd}}$ February 2019, respectively. The night-to-night variations are significantly reduced compared to the solutions obtained with a free cavity width fit, with the redder orders driving most of the remaining variability.

\subsection{Intra-night stability}

\begin{figure}[htb]
	\centering
	\includegraphics[width=1\linewidth]{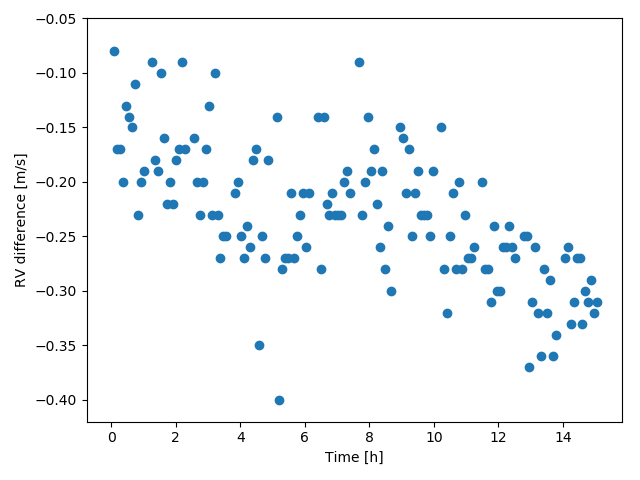}  
    \includegraphics[width=1\linewidth]{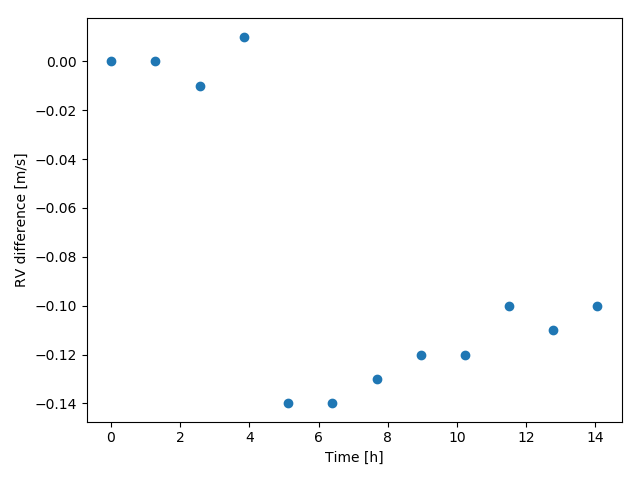}      
	\caption[RV variation over time]{Radial velocity variations over time for a CCF RV of an FP frame (taking it as an artificial 'star'), computed using wavelength solutions with a fixed HC frame and varying FP frames (top), or varying HC frames and a fixed FP frame (bottom). The RVs have been drift-corrected.}
	\label{fig: RV-intra-night}
\end{figure}

To verify the intra-night stability of the HC-FP wavelength solution, we computed solutions using sequences of 1 HC and 10 FP frames taken continuously over a 14h period. We adopted two test configurations: fixing the HC frame and varying the FP frame, and fixing the FP frame and varying the HC frame. For each of these, we obtained HC-FP wavelength solutions using method FP2. We then computed the RV of an FP frame (taking it as an artificial 'star') using the different wavelength solutions. The resulting RV variations (corrected for the spectrograph drift by using the FP calibration fibre CCF, as described in Section \ref{FP-sol-desc}) are shown in Fig. \ref{fig: RV-intra-night}. For the case of a fixed HC frame and varying FP frames, the RV variations show a slow downward trend, which is probably due to the intrinsic drift of the FP étalon with respect to the absolute wavelength reference. For the case of a fixed FP frame and varying HC frames, the RV variations are very small, with an amplitude comparable to the photon noise.

\subsection{Impact on RV error}\label{ch2: RV-err-imp}

Ultimately, the interest of an accurate and stable wavelength solution for a spectrograph lies in being able to measure precise RVs. In the SPIRou DRS, RVs are measured by the CCF method - that is, by cross-correlating a binary stellar mask with the observed spectrum. This cross-correlation is first performed order by order; these are then summed together, and a Gaussian fit is performed, the final RV being measured as the centre of the Gaussian. The RVs per order and the combined RV are all stored. If the star is observed with simultaneous FP on the calibration fibre, the instrumental drift is computed by the same recipe, cross-correlating the FP spectrum to a binary FP mask.

To evaluate the impact of the wavelength solution on the RVs, we selected observations of two stars, which we shall refer to as star A and star B, chosen as they are the brightest targets observed close to the centre of the February 2019 run. For each star, we computed the RVs changing the input wavelength solution. To generate the wavelength solutions, we adopted method FP2 and a fixed cavity width fit as this was shown to provide the most accurate and stable set of wavelength solutions. We used the CCF computation from the SPIRou DRS to obtain the RVs. Both stars were observed with simultaneous FP calibration, so the drift was also computed from the FP CCFs, as described in Section \ref{FP-sol-desc}. The results are summarised in Table \ref{tab: CCF-RV-res}. The standard deviation of the differences from the RV obtained for 13 February (taken as the reference point) is $\mathrm{0.67\,m\,s^{-1}}$ for star A, and $\mathrm{0.40\,m\,s^{-1}}$ for star B. We observed and corrected for large drifts (median value $\mathrm{11.5\,m\,s^{-1}}$); these drifts are induced by the fact we have fixed a single set of prior calibrations for extracting all the files, which creates offsets between the wavelength solutions, as described in Sect. \ref{ch2:prev-calib-impact}. The physical origin of these day-to-day drifts may lie in many different factors: The instrument is known to be sensitive to vibrations; the room in which the FP is located is not stabilised; there are known jumps at each thermal cycle of the cryostat; among others.

\begin{figure*}[htb]
	\centering
	\begin{tabular}[c]{cc}
    		\includegraphics[width=0.49\linewidth]{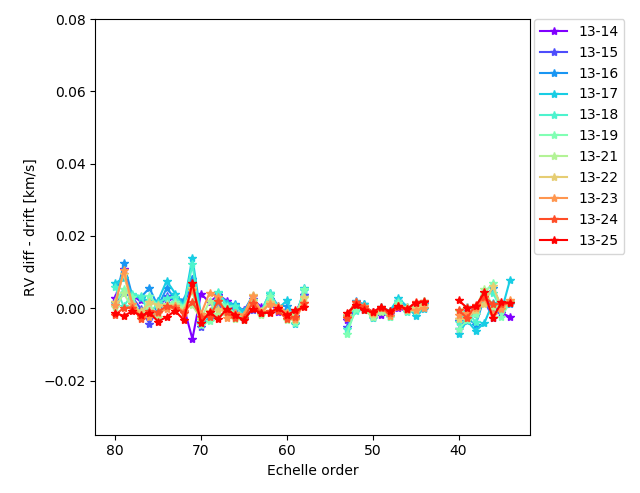}  &
      		\includegraphics[width=0.49\linewidth]{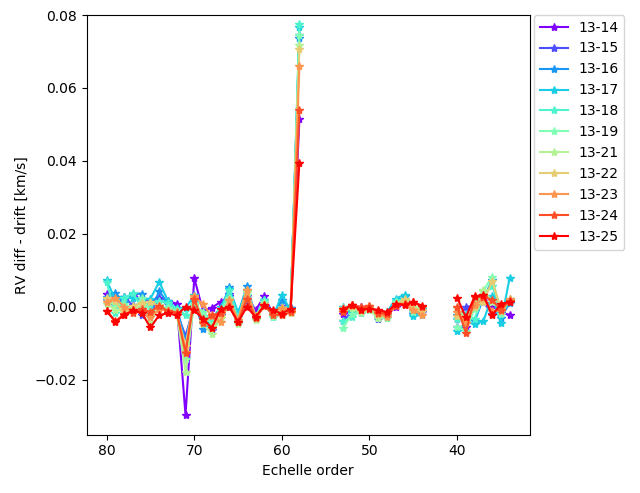}   \\
    \end{tabular} 
	\caption[CCF RV differences per order]{Drift-corrected CCF RV differences per order for star A (left) and star B (right), using wavelength solutions from different nights. The strong variations of echelle order 58 for star B are due to a poor CCF fit.}
	\label{fig:CCF-RV-ord}
\end{figure*}

\begin{table*}
\centering
    \caption[Summary of CCF RV differences.]{Summary of CCF RV differences using different wavelength solutions.}  
    \begin{tabular}{rrrr}
    \hline \hline
  Star & Wavelength  & RV diff (all  & RV diff (sel.    \\
   & sol. night & orders) [$\mathrm{m\,s^{-1}}$] & orders) [$\mathrm{m\,s^{-1}}$] \\ 
  \hline
  star A   & 13 Feb  & ---           & ---     \\
  &  14 Feb  & 0.89       & 0.89       \\
  &  15 Feb  & 0.20        & 0.20         \\
  &  16 Feb  & 0.22       & 0.22       \\
  &  17 Feb  & 0.68       & 0.68       \\
  &  18 Feb  & 0.26       & 0.25       \\
  &  19 Feb  & 0.36       & 0.35       \\
  &  21 Feb  & -0.39      & -0.40        \\
  &  22 Feb  & 0.07       & 0.07       \\
  &  23 Feb  & 0.09       & 0.09       \\
  &  24 Feb  & -0.54      & -0.55      \\
  &  25 Feb  & 1.93       & -0.78      \\
  \cline{2-4}
  & $\sigma_{RV diff}$ & 0.67 & 0.50 \\
  \hline
 star B & 13 Feb & ---           &     --- \\           
 & 14 Feb  & 1.01       &    1.01 \\          
 & 15 Feb  & 0.20        &    0.23 \\       
 & 16 Feb  & 0.07       &   0.10 \\          
 & 17 Feb  & -0.61      &    -0.58 \\          
 & 18 Feb  & 0.14       &   0.17 \\         
 & 19 Feb  & 0.39       &   0.42 \\           
 & 21 Feb  & 0.18       &    0.19 \\        
 & 22 Feb  & 0.08       &   0.10 \\          
 & 23 Feb  & 0.22       &   0.23 \\          
 & 24 Feb  & -0.09      &   -0.08 \\         
 & 25 Feb  & -0.20       &    -0.20 \\  
   \cline{2-4}
  & $\sigma_{RV diff}$ & 0.50 & 0.39 \\  
    \hline \\
    \end{tabular}
    \label{tab: CCF-RV-res}
\end{table*}
Although the overall variations are small, it is worthwhile to inspect the CCFs in more detail. Fig. \ref{fig:CCF-RV-ord} shows the drift-corrected difference in CCF RVs per order for each star. Gaps correspond to orders for which no CCF could be calculated (generally due to a very low or null atmospheric transmission). It is clear that some orders are far more variable than others, and may be driving the RV differences. It is hard to say whether this is due to solely to the wavelength solution (since, for instance, Fig. \ref{fig:FP-night-night} does not show significantly higher night-to-night variability in these orders), or whether there are also factors due to the CCF at play (e.g. smaller spectral lines in these orders whose fit could be more impacted by small shifts in wavelength solution). 

In particular, for star A the RVs for echelle orders 71 (central wavelength 1097 nm) and 79 (central wavelength 984 nm) have standard deviations of $\mathrm{\sim 5\,m\,s^{-1}}$, while the rest are below $\mathrm{\sim 3\,m\,s^{-1}}$. Likewise, for star B, the RVs for echelle orders 58 (central wavelength 1348 nm) and 71 have standard deviations of $\mathrm{\sim 12\,m\,s^{-1}}$ and $\mathrm{\sim 8\,m\,s^{-1}}$ respectively, while the rest are below $\mathrm{\sim 3\,m\,s^{-1}}$. For order 58 for star B, in particular, closer analysis shows the Gaussian fit to the CCF is clearly poor, explaining the high RV offset and dispersion. Recalculating the RVs excluding the worst two orders for each star, the differences are generally slightly reduced (last column of Table \ref{tab: CCF-RV-res}), with a standard deviation of $\mathrm{0.50\,m\,s^{-1}}$ for star A and $\mathrm{0.39\,m\,s^{-1}}$ for star B, though the impact is not large. 

There are many factors that may contribute to the remaining noise. The photon noise is at the $\mathrm{\leq 0.10\,m\,s^{-1}}$ level, so is unlikely to be a major contributor. The spectrograph drift correction may be playing a role (over 12 days, the drift is of the order of $\mathrm{\sim 50\,m\,s^{-1}}$. In standard operations with daytime calibrations, these drifts would not appear; as we fixed a single set of calibrations, they need to be removed separately). For the wavelength solution, inaccuracies in the UNe catalogues and fit instabilities may increase the noise. Several detector-related effects may also be contributing, such as cosmic rays, bad pixels, the known persistence on H4RG detectors, or intra-pixel response variations.

\section{Conclusions and future perspectives} \label{s: Conc}

We have implemented and tested different methods of generating a wavelength-pixel correspondence, using either HC lamps alone or the combination of HC lamps with an FP étalon. The HC lamps alone did not provide sufficient accuracy, being at the level of $\mathrm{\sim 2 {\text -}3\,m\,s^{-1}}$ internal error, while the error budget prevision was of $\mathrm{<0.45\,m\,s^{-1}}$. The combined HC-FP solutions, on the other hand, have an excellent internal error of $\mathrm{\sim 0.15\,m\,s^{-1}}$. The stability from one night to the next is complicated by the dependence of the wavelength solution on the previous calibrations, especially on the slit determination. Fixing all prior calibrations produces a noticeable ($\mathrm{\sim 10 {\text -}50\,m\,s^{-1}}$) but constant RV offset between solutions; when this offset is removed, the night-to-night variations are greatly diminished. We analysed the impact of changing the wavelength solution on the RV calculations, finding that the calculated RVs remain fairly consistent with $\mathrm{\sim 0.5 {\text -}0.7\,m\,s^{-1}}$ dispersions, and that the drift computation is efficient at removing the RV offset between wavelength solutions computed with a fixed set of previous calibrations.

This article is based primarily on the current stable version of the DRS, 0.5.000. Significant changes have been planned for future versions, which the DRS team is currently working on. Several of these changes are either directly on the wavelength solution, or are expected to impact it, such as the implementation of a set of 'master' calibrations that are not expected to change on a nightly basis, but only per run or even per thermal cycle. One of these master calibrations will be a master wavelength solution, with nightly calibrations measuring only the offset from this master solution. Another problem to be dealt with is potential variations between solutions obtained for the science and calibration fibres. In version 0.5.000, these are computed independently, which can lead to cases such as the wavelength solution failing quality controls for one fibre but not another. Upcoming versions will anchor the calibrations together to avoid this problem.

Wavelength solutions combining HC and FP exposures are implemented for CARMENES \citep{Bauer15, Caballero16} and have recently been tested for HARPS \citep{Cersullo19}. In both cases, these wavelength solutions are shown to be suitable for reaching a $\mathrm{1\,m\,s^{-1}}$ overall RV precision. We anticipate that this will also be the case for SPIRou.

\begin{acknowledgements}
The authors wish to recognise and acknowledge the very significant cultural role and reverence that the summit of Maunakea has always had within the indigenous Hawaiian community.  We are most fortunate to have the opportunity to conduct observations from this mountain.
This work was supported by the Programme National de Planétologie (PNP) of CNRS/INSU, co-funded by CNES. We acknowledge funding from ANR of France under contract number ANR-18-CE31-0019 (SPlaSH).
This research made use of matplotlib, a Python library for publication quality graphics \citep{Hunter:2007}; SciPy \citep{jones_scipy_2001}; IPython package \citep{PER-GRA:2007}; Astropy, a community-developed core Python package for Astronomy \citep{2018AJ....156..123A, 2013A&A...558A..33A}; NumPy \citep{van2011numpy}; Astroquery \citep{2019AJ....157...98G}; ds9, a tool for data visualisation supported by the Chandra X-ray Science Center (CXC) and the High Energy Astrophysics Science Archive Center (HEASARC) with support from the JWST Mission office at the Space Telescope Science Institute for 3D visualisation. 
We thank all SPIRou partners for their funding contributions to the SPIRou project, whose construction cost (including reviews
and travels) reached a total of $\cong$5Me, namely the IDEX initiative at UFTMP, UPS, the DIM-ACAV programme in Region Ile de France, the MIDEX initiative at AMU, the Labex@OSUG2020 programme, UGA, INSU/CNRS, CFI, CFHT, LNA, CAUP and DIAS. We are also grateful for generous amounts of in-kind manpower allocated to SPIRou by OMP/IRAP, OHP/LAM, IPAG, CFHT, NRC-H, UdeM, UL, OG, LNA and ASIAA, amounting to a total of about 75 FTEs including installation and ongoing upgrades.
This work has been carried out within the framework of the National Centre of Competence in Research PlanetS supported by the Swiss National Science Foundation.
\end{acknowledgements}

\bibliographystyle{aa} % style aa.bst
\bibliography{biblio} % your references Yourfile.bib

\begin{appendix}\label{appendix}
\section{Updated wavelength catalogue}

In this appendix, we present our updated list of UNe lines, as described in Sect. \ref{s:Perf}. The original wavelength values come from the catalogues of \cite{Redman11} or \cite{Sarmiento18}. Only the first few lines are shown; the full catalogue is available in electronic form at the CDS.

\begin{table}[htb]
    \caption{Updated list of UNe line wavelengths}
    \begin{tabular}{ll}
        \hline\hline
        Original wavelength [nm] & Updated wavelength [nm] \\
        \hline 
        965.3596 & 965.35960878\\
        965.5902 & 965.59063720\\
        965.9620 & 965.96233723\\
        967.0700 & 967.07066522\\
        967.1665 & 967.16724282\\
        $\cdots$ & $\cdots$\\
        \hline
    \end{tabular}
\end{table}

\end{appendix}

\end{document}